\documentclass{article}
\usepackage{spconf,amsmath,graphicx}
\usepackage{cite}
\usepackage{mathtools}
\usepackage{multicol}
\usepackage{multirow}
\usepackage{subfig}
\usepackage{threeparttable}

\title{Detecting Mismatch between Text Script and Voice-over Using Utterance Verification Based on Phoneme Recognition Ranking}
\name{Yoonjae Jeong, Hoon-Young Cho}
\address{
    Speech Lab, AI Center, NCSOFT, Seongnam, Gyeonggi-do, Republic of Korea \\
    \textit{\{yjeong, hycho\}@ncsoft.com}
}
\begin{document}
%

\maketitle

\let\thefootnote\relax\footnotetext {
Copyright 2020 IEEE. Published in the IEEE 2020 International Conference on Acoustics, Speech, and Signal Processing (ICASSP 2020), scheduled for 4-9 May, 2020, in Barcelona, Spain. Personal use of this material is permitted. However, permission to reprint/republish this material for advertising or promotional purposes or for creating new collective works for resale or redistribution to servers or lists, or to reuse any copyrighted component of this work in other works, must be obtained from the IEEE. Contact: Manager, Copyrights and Permissions / IEEE Service Center / 445 Hoes Lane / P.O. Box 1331 / Piscataway, NJ 08855-1331, USA. Telephone: + Intl. 908-562-3966.
}

\begin{abstract}
    The purpose of this study is to detect the mismatch between text script and voice-over. For this, we present a novel utterance verification (UV) method, which calculates the degree of correspondence between a voice-over and the phoneme sequence of a script. We found that the phoneme recognition probabilities of exaggerated voice-overs decrease compared to ordinary utterances, but their rankings do not demonstrate any significant change. The proposed method, therefore, uses the recognition ranking of each phoneme segment corresponding to a phoneme sequence for measuring the confidence of a voice-over utterance for its corresponding script. The experimental results show that the proposed UV method outperforms a state-of-the-art approach using cross modal attention used for detecting mismatch between speech and transcription.
\end{abstract}
\section{Introduction}
The popularization of multimedia and computer games has enabled the emergence of a large number of text scripts and their voice-overs. For example, in massively multiplayer online role-playing games (MMORPGs), non-player characters (NPCs) deliver their messages to users through captions and voice-overs. In several cases, game developers outsource the voice-overs of NPCs' text scripts to voice actors and manually verify if the outcomes match to their corresponding scripts. This type of a manual verification is extremely time-consuming and labor-intensive; therefore, there is a growing need to automate the verification process.

For this, the application of automatic speech recognition (ASR) system is not considered to be effective, because these systems commonly produce the incorrect output for the utterances containing out-of-vocabulary (OOV) words. Moreover, the voice-overs in a game domain contain more OOVs than ordinary speech.

Utterance verification (UV) \cite{doi:10.1016/J.SPECOM.2004.12.004} is one of the key technologies that can be used to deal with this problem. As a component of ASR systems, the UV prevents the ASR system from producing incorrect recognition results by evaluating the confidence between a user's utterance and its recognized text. The UV technique can be applied to detect a mismatch between a script and its voice-over. The \textit{utterance verifier} receives a pair of a script and its voice-over (text and waveform) as an input. Subsequently, with the use of an acoustic model, the \textit{verifier} calculates the degree of correspondence between them.

Most of the acoustic models are trained using a read- or natural-style speech database because they are relatively easy to construct, and there are several public corpora. However, the speech style in a computer game domain is entirely different. The voice actors tend to utter with exaggerated and emotional intonation.

We found that the conventional UV algorithms are not suitable for verifying the exaggerated or emotional utterances when an acoustic model trained based on a read- and natural-style speech database is used. The conventional UV algorithm measures the confidence based on the gap of phoneme recognition probabilities between an acoustic model and its anti-phoneme model. However, the differences in the speech style reduce the recognition probabilities of exaggerated utterances, and therefore, it induces the decrease of the probability gap.

To resolve the aforementioned problem, we devise a novel UV algorithm based on phonemic recognition rankings. It is observed that the phoneme recognition rankings do not significantly change regardless of the speech styles. The proposed UV algorithm calculates the average phoneme recognition ranking of each speech segment of a phoneme sequence corresponding to its text script as the confidence measure.

\section{Related Work}
Utterance verification (UV) formulates the confidence measuring problem as a statistical hypothesis testing of the \textit{null} hypothesis against the \textit{alternative} hypothesis \cite{doi:10.1109/89.568733, doi:10.1109/ICASSP.1995.479528, doi:10.1109/89.544527}. The \textit{null} hypothesis is that, if a speech recognizer recognizes an utterance as a correct word (or phoneme) sequence, the unit word (or phoneme) of the sequence is also correctly recognized in its corresponding speech segment. Whereas, the \textit{alternative} hypothesis is that the words (or phonemes) are wrongly recognized and therefore cannot originate from their speech segments.

The studies concerned with the prediction of automatic speech recognition (ASR) errors have focused on building classifiers for discovering incorrect outcomes of an ASR system for given utterances \cite{Fayolle2010, doi:10.1109/ICASSP.2012.6288880, doi:10.1109/ICASSP.2013.6639103,  doi:10.1109/ICASSP.2014.6854012, Korenevsky2015, doi:10.1109/ICASSP.2015.7178796, doi:10.1109/AICCSA.2016.7945669, doi:10.1016/J.SPECOM.2017.02.009, doi:10.1109/SLT.2018.8639602}. The classifiers use the combined features generated from various sources, including the intermediate results of the decoding process, and usually outperform the UV approaches. However, because these studies aim to determine the accuracy of the ASR result for an input speech, it does not satisfy the goal of the present study.

Huang and Hain \cite{doi:10.21437/Interspeech.2019-2125} recently proposed a new method for detecting a mismatch between speech and transcription using a cross-modal attention mechanism. They also tried to present a robust method that can be applied even in the absence of a well-defined lexicon and large corpus. Because their objective is similar to ours, we use \cite{doi:10.21437/Interspeech.2019-2125} as one of the baselines of our experiments.

\section{Proposed Method}
The procedure followed by the proposed utterance verification (UV) system is described below. (1) The system extracts the features from a given voice-over utterance. The features consist of the 13-dimensional Mel-Frequency Cepstral Coefficients (MFCC) appending delta and delta-delta values. (2) We generate the possible phoneme sequences from a given text script using the pronunciation dictionary created based on the grapheme-to-phoneme (G2P) method. (3) The forced alignment module finds the best phoneme sequence of the script and its voice-over and aligns the phoneme sequence to the speech features. (4) Next, our system runs the likelihood ratio test, through which the overly mismatched pair of script and voice-over are filtered. (5) In the final stage, the proposed average phoneme ranking (APR) based UV validates the voice-overs that qualified in the previous stage. The system then measures the correspondence between each script and its voice-over as the average of the phonemic recognition rankings of each phoneme in the aligned phoneme sequence. The higher the average ranking, the greater is the confidence match between a script and its voice-over.

\subsection{Likelihood Ratio Test (LRT) based Utterance Verification}
The likelihood ratio test (LRT) has widely been the basis for numerous utterance verification (UV) methods. The LRT is a statistical test that determines the goodness of a \textit{null} model against an \textit{alternative} model called the anti-model. \cite{doi:10.1109/89.568733} is adopted as our baseline LRT-based UV method because it is a representative LRT-based UV study and has also been the basis for several subsequent studies.

Equation (\ref{eqn:llr}) represents the log-likelihood ratio (LLR) between a script $t$ and voice-over $u$. $H_0$ is a model representing the probability that an acoustic model recognizes $u$ as the $t$, and the $H_1$ is an anti-model of $H_0$. The $g(t, u)$ and $G(t, u)$ represent the log-likelihoods of $H_0$ and $H_1$, respectively. If LLR is higher than threshold $\tau$, then $u$ is considered to match $t$.
\begin{equation}
    LLR(t, u) = \mathrm{log} \frac{P(H_0)}{P(H_1)} = g(t, u) - G(t, u) > \tau
    \label{eqn:llr}
\end{equation}
        
\subsection{Average Phoneme Ranking (APR) based Utterance Verification}
\begin{figure}[t]
    \centering
    \subfloat[Read Style (LLR)]{
        \centering
        \includegraphics{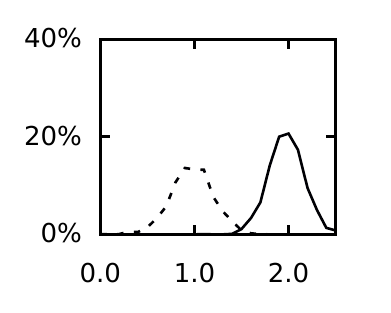}
        \label{fig:dist-llr.read}
    }
    \subfloat[Online Game (LLR)]{
        \centering
        \includegraphics{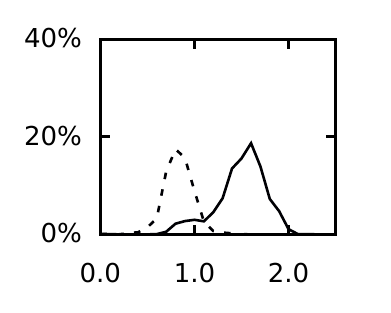}
        \label{fig:dist-llr.game}
    }
    \newline \noindent
    \subfloat[Read Style (APR)]{
        \centering
        \includegraphics{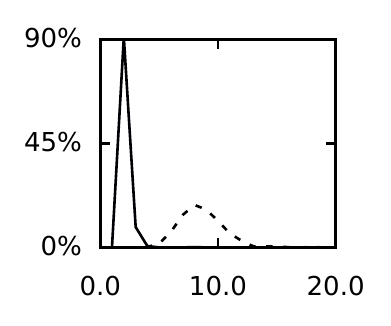}
        \label{fig:dist-apr.read}
    }
    \subfloat[Online Game (APR)]{
        \centering
        \includegraphics{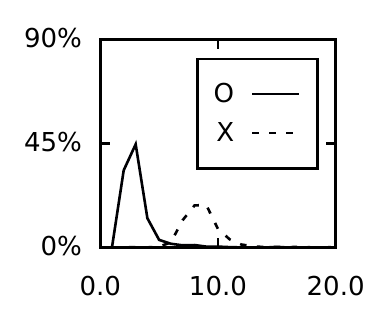}
        \label{fig:dist-apr.game}
    }
    \caption{Distribution of the script-voiceover pairs on LLR and APR. The solid- and dashed-line indicate the correct (O) and incorrect (X) pairs, respectively.}    
    \label{fig:dist}
\end{figure}

When we apply the LRT-based utterance verification (UV) to the exaggerated voice-overs, significant performance degradation occurs. Figure \ref{fig:dist} shows the distribution of the correct script-voiceover pairs (O) and the incorrect ones (X) on the log-likelihood ratio (LLR) and the APR.

For the read speech, which has the same speech style as the training data of the acoustic model, the LRT-based UV clearly separates the correct and incorrect pairs at a threshold of 1.5 (Figure \ref{fig:dist-llr.read}). However, when we apply the threshold to the exaggerated speech style of the online game, a significant number of correct pairs are determined to be incorrect. Moreover, the distribution of the correct and incorrect pairs significantly overlap (Figure \ref{fig:dist-llr.game}). This is because the recognition probabilities decrease due to the different speech styles of the acoustic model. In contrast, the recognition rankings of phonemes do not change significantly, regardless of the speech styles, as shown in Figure \ref{fig:dist-apr.read} and \ref{fig:dist-apr.game}. Only a slight drop in ranking occurs in the correct exaggerated voice-over.

Based on this observation, we measure the correspondence between a script $t$ and voice-over $u$ as the average ranking of phoneme recognition in Equation (\ref{eqn:apr}). We call this new UV method as the average phoneme ranking (APR) based UV. In the equation below, $p_i$ indicates the $i$-th phoneme of the script $t$, and $f_i$ is the speech features corresponding to the $i$-th phoneme. $rank(p_i, f_i)$ is the phoneme recognition rank of $p_i$ for $f_i$. If the APR is less than threshold $\theta$, then $u$ is considered to correspond to $t$. $N$ is the number of phonemes in the phoneme sequence.
\begin{equation}
    APR(t, u) = \frac{1}{N} \sum_{i=1}^{N}{rank(p_i, f_i)} < \theta
    \label{eqn:apr}
\end{equation}

\subsection{Two-stage APR-based Utterance Verification}
We identified some rare cases where the phoneme recognition rankings are high, although the overall recognition probabilities are extremely low. To avoid the occurrence of such scenarios, we devise a two-stage APR-based UV. Equation (\ref{eqn:apr_2_stage}) shows the two-stage verification method that combines the LRT-based UV and the APR-based UV.
\begin{align}
    APR_{\mathrm{2-stage}}(t, u) & =
        \begin{dcases}
            | P |     & \text{if } LLR(t, u) \leq \tau \\
            APR(t, u) & \text{otherwise}
        \end{dcases} \nonumber \\
    APR_{\mathrm{2-stage}}(t, u) & < \theta \label{eqn:apr_2_stage}
\end{align}

If the LLR value of the LRT-based UV is less than or equal to threshold $\tau$, the lowest phoneme recognition ranking value ($|P|$) is assigned. Otherwise, the average phoneme ranking calculated via the APR-based UV is used to measure the correspondence between script $t$ and voice-over $u$.

\section{Experiment}
\subsection{Experimental Setup}
\subsubsection{Acoustic Model}
We build a Deep Neural Network (DNN) based acoustic model and a GMM model for our experiment. The former is used for the forced alignment process, and the latter is used for scoring a segmented speech in a phoneme unit. Although the GMM model is a rudimentary acoustic model, it demonstrates sufficiently impressive performance for verifying general pronunciation. Moreover, in our test environment, the GMM model performs better than the DNN model for phoneme scoring. Using Kaldi-ASR toolkits \cite{Povey2011}, we train the English and Korean acoustic models based on the LibriSpeech ASR corpus \cite{doi:10.1109/ICASSP.2015.7178964} and roughly 793,000 read-style Korean utterances, respectively.

\subsubsection{Test Sets}
Considering the purpose of the experiment, we employ two types of test sets. The first set is concerned with the comparison of the proposed method with those of previous studies. The test set of the WSJ-CAM0 corpus \cite{Hain2013} is adopted to compare our method with that of Huang and Hain's study \cite{doi:10.21437/Interspeech.2019-2125}, which is the state-of-the-art for detecting a mismatch between speech and transcription. To create the mismatched samples, we randomly delete, insert, and substitute four words in the original test data. The test data for evaluation contains 331 matched samples and 331 mismatched ones.

The second type of test set is used to detect a mismatch between text script and voice-over. We create three test sets (DICT01, BNS-1, and BNS-2). DICT01 is the 1,600 read-style utterances excerpted from a Korean speech DB (DICT01) \cite{SiTEC}, and BNS-1 comprises the pairs of voice-over and script randomly selected from an MMORPG game of NCSOFT \cite{NCSOFT}. BNS-2 comprises 483 voice-overs with various tones and sound effects. To create the mismatched pairs, we assign an arbitrary text script to a voice-over similar to the real-world scenario. As a result, we obtain the balanced test sets of the correct and incorrect pairs, as presented in Table \ref{tab:test_sets}.

\begin{table}[ht]
    \caption{Excerpted test sets from a speech database and an MMORPG}
    \label{tab:test_sets}
    \centering
    \begin{tabular}{l|l|r|r}
        \hline
        Test Set & Description & Correct & Incorrect \\
        \hline
        DICT01 & Read-style                 & $1,600$ & $1,600$ \\
        BNS-1  & Exaggerated-style          & $1,600$ & $1,600$ \\
        BNS-2  & + various tones \& effects & $483$   & $483$   \\
        \hline
    \end{tabular}
\end{table}

\subsection{Evaluation}
\subsubsection{Comparison with Previous Work}
We first evaluate the proposed APR-based utterance verification (UV) by comparing it against previous work, as listed in Table \ref{tab:comparision_with_previous_work}. We chose the cross-modal attention of Huang and Hain \cite{doi:10.21437/Interspeech.2019-2125} and the LRT-based UV \cite{doi:10.1109/89.568733} as baselines. \cite{doi:10.21437/Interspeech.2019-2125} is the state-of-the-art based on a deep-learning approach and is used for detecting a mismatch between speech and transcription, and \cite{doi:10.1109/89.568733} is the conventionally representative UV method. We build the test set from the WSJ-CAM0 corpus in the same way as \cite{doi:10.21437/Interspeech.2019-2125}.

\begin{table}[ht]
    \caption{Comparison of 4-word mismatch detection accuracy for deletion (Del), insertion (Ins), and substitutions (Sub) in the WSJ-CAM0 test set}
    \label{tab:comparision_with_previous_work}
    \centering
    \begin{tabular}{l|c|c|c||c}
        \hline
            & Del & Ins & Sub & Avg \\
        \hline
        Huang \& Hain \cite{doi:10.21437/Interspeech.2019-2125}
            & 0.781 & 0.792 & 0.558 & 0.710 \\
        LRT \cite{doi:10.1109/89.568733}
            & 0.605 & 0.798 & 0.670 & 0.691 \\
        \hline
        Proposed APR
            & 0.730 & 0.986 & 0.918 & 0.878 \\
        $\text{Proposed APR}_{\text{2-stage}}$
            & 0.731 & 0.986 & 0.920 & 0.879 \\
        \hline
    \end{tabular}
\end{table}

Overall, our proposed APR-based UV shows an increase in accuracy by approximately 0.169 (23.8\%) and 0.188 (27.2\%) compared to \cite{doi:10.21437/Interspeech.2019-2125} and \cite{doi:10.1109/89.568733}, respectively. Our method shows improvements of approximately 0.194 (24.5\%) and 0.362 (64.9\%) in terms of accuracy for insertion (Ins) and substitution errors (Sub) compared to \cite{doi:10.21437/Interspeech.2019-2125}. In the deletion errors (Del), there was a performance decrease of  approximately 0.05, but this decrease is much smaller than the improvement of the Ins and Sub errors.

\subsubsection{Performances for Detecting Mismatch between Text Script and Voice-over}
In the test sets of text scripts and voice-overs, we perform an experiment for comparing the performances of the proposed APR-based UV and the conventional LRT-based UV. APR is an alternative to the log-likelihood ratio (LLR) of the LRT-based UV. For the evaluation, we apply the optimized thresholds (i.e., $\tau$ and $\theta$) that show the best performances for test sets to each method. Table \ref{tab:comparison_in_voiceover} presents the results.

\begin{table}[ht]
    \caption{Comparison between the proposed APR-based UV and the conventional LRT-based UV with the optimized thresholds.}
    \label{tab:comparison_in_voiceover}
    \centering
    \begin{threeparttable}
        \begin{tabular}{l|c|c|c|c|c}
            \hline
            \multirow{2}{*}{Test Set} &
            \multicolumn{2}{c|}{LRT} &
            \multicolumn{3}{c}{APR} \\
            &
            \multicolumn{1}{c|}{ACC} &
            \multicolumn{1}{c|}{$\tau$} &
            \multicolumn{1}{c|}{ACC} &
            \multicolumn{1}{c|}{$\theta$} &
            \multicolumn{1}{c}{$\Delta$} \\
            \hline
            DICT01 & 0.992 & 1.5 & 0.998 & 4.0 & +0.006 (0.6\%) \\
            BNS-1  & 0.930 & 1.2 & 0.968 & 5.0 & +0.038 (4.1\%) \\
            BNS-2  & 0.901 & 1.1 & 0.959 & 6.0 & +0.058 (6.4\%) \\
            \hline
        \end{tabular}
        \begin{tablenotes}
            \small
            \item[1] ACC is the accuracy, and $\Delta$ is the accuracy improvement of the APR-based UV compared to the LRT-based UV
        \end{tablenotes}
    \end{threeparttable}
\end{table}

For read-style utterances (DICT01), the APR-based UV obtains a performance gain of approximately 0.006 (0.6\%) in terms of accuracy. The improvement in the DICT01 is not significant, but the APR-based UV presents a significant improvement in exaggerated voice-overs. For BNS-1 and BNS-2, our method shows an improvement of 0.038 (4.1\%) and 0.058 (6.4\%), respectively. As the improvement is more in BNS-2, the proposed APR-based UV appears to be more robust for a variety of tones and sound effects as compared to the LRT-based UV.

\subsubsection{Robustness to Threshold}
To inspect the robustness of threshold values to datasets, we investigated the performance drops for the exaggerated voice-overs when the threshold ${\tau} = 1.2$ and ${\theta} = 4.0$ optimized for read-speech utterances are applied. As detailed in Table \ref{tab:robustness}, the declines in the performance of the LRT-based UV in BNS-1 and BNS-2 are -0.117 (-14.4\%) and -0.228 (-33.8\%), respectively. However, those in the case of the APR-based UV are -0.016 (-1.7\%) and -0.059 (-6.6\%), respectively, which are remarkably lower than that of the LRT-based UV.

\begin{table}[ht]
    \caption{Performance degradation in the exaggerated voice-overs, when applying the optimized thresholds of the read-speech utterances.}
    \label{tab:robustness}
    \centering
    \begin{threeparttable}
        \begin{tabular}{l|c|c|c|c}
            \hline
            \multirow{2}{*}{Test Set} &
            \multicolumn{2}{c|}{LRT} &
            \multicolumn{2}{c}{APR} \\
            &
            \multicolumn{1}{c|}{ACC} &
            \multicolumn{1}{c|}{$\Delta$} &
            \multicolumn{1}{c|}{ACC} &
            \multicolumn{1}{c}{$\Delta$} \\
            \hline
            \multirow{2}{*}{BNS-1} & \multirow{2}{*}{0.813} & -0.117    & \multirow{2}{*}{0.952} & -0.016 \\
                                   &                        & (-14.4\%) &                   & (-1.7\%) \\
            \hline
            \multirow{2}{*}{BNS-2} & \multirow{2}{*}{0.674} & -0.228    & \multirow{2}{*}{0.900} & -0.059 \\
                                   &                        & (-33.8\%) &                   & (-6.6\%) \\
            \hline
        \end{tabular}
        \begin{tablenotes}
            \small
            \item[1] $\Delta$ indicates the reduction in accuracy from the optimized threshold for the test set.
        \end{tablenotes}
    \end{threeparttable}
\end{table}

\subsubsection{Effects of Two-stage Approach}
We finally investigate the effect of the proposed two-stage APR-based UV. Table \ref{tab:peformance_improvement_of_two-stage_apr} presents its improvement in comparison with a pure APR-based UV. Although the improvement is minimal, the two-stage APR-based UV compensates for a few errors of the pure APR-based UV.

\begin{table}[ht]
    \caption{Performance improvement of the two-stage APR-based UV.}
    \label{tab:peformance_improvement_of_two-stage_apr}
    \centering
    \begin{tabular}{l|c|c|c}
        \hline
        Test Set & APR & $\text{APR}_{\text{2-stage}}$ & $\Delta$ \\
        \hline
        BNS-1 & 0.9675 & 0.9677 & +0.0002 \\
        BNS-2 & 0.9592 & 0.9598 & +0.0006 \\
        \hline
    \end{tabular}
\end{table}

\section{Conclusions and Future Work}
In this paper, an APR based utterance verification (UV) method was proposed. Our experimental results showed that the proposed method showed performance improvements over the state-of-the-art as well as the conventional LRT-based UV when detecting mismatches between speech and transcriptions. Additionally, our method showed only a small amount of performance degradation with exaggerated voice-overs, even though the model was optimized to read-style utterances.

Although the proposed method showed encouraging results compared to previous approaches, it still has some limitations. One such limitation is concerned with the handling of deletion errors. The proposed APR-based UV method showed performance degradation for speech and transcript pairs with missing words when compared to that of the state-of-the-art. The other limitation is concerned with the handling of laughing-style utterances. Since laughing-style utterances are pronounced differently depending on the situation, transcribing them to proper phoneme sequences is a challenging task. As a direction for future research, we are currently working on the two aforementioned issues.

\vfill\pagebreak

\bibliographystyle{IEEEbib}
\bibliography{strings,refs}

\begin{thebibliography}{10}

\bibitem{doi:10.1016/J.SPECOM.2004.12.004}
Hui Jiang,
\newblock ``{Confidence measures for speech recognition: A survey},''
\newblock {\em Speech Communication}, vol. 45, no. 4, pp. 455--470, 2005.

\bibitem{doi:10.1109/89.568733}
Mazin~G. Rahim, Chin-Hui Lee, and Biing-Hwang Juang,
\newblock ``{Discriminative utterance verification for connected digits
  recognition},''
\newblock {\em IEEE Transactions on Speech and Audio Processing}, vol. 5, no.
  3, pp. 266--277, 1997.

\bibitem{doi:10.1109/ICASSP.1995.479528}
Richard~C. Rose, Biing-Hwang Juang, and Chin-Hui Lee,
\newblock ``{A training procedure for verifying string hypotheses in continuous
  speech recognition},''
\newblock in {\em Proceedings of the 1995 International Conference on
  Acoustics, Speech, and Signal Processing (ICASSP)}, 1995, pp. 281--284.

\bibitem{doi:10.1109/89.544527}
Rafid~A. Sukka and Chin-Hui Lee,
\newblock ``{Vocabulary independent discriminative utterance verification for
  nonkeyword rejection in subword based speech recognition},''
\newblock {\em IEEE Transactions on Speech and Audio Processing}, vol. 4, no.
  6, pp. 420--429, 1996.

\bibitem{Fayolle2010}
Julien Fayolle, Fabienne Moreau, Christian Raymond, Guillaume Gravier, and
  Patrick Gros,
\newblock ``{CRF-based Combination of Contextual Features to Improve A
  Posteriori Word-level Confidence Measures},''
\newblock in {\em Proceedings of Interspeech 2010}, 2010, pp. 1942--1945.

\bibitem{doi:10.1109/ICASSP.2012.6288880}
Matthew Gibson and Thomas Hain,
\newblock ``{Application of SVM-based correctness predictions to unsupervised
  discriminative speaker adaptation},''
\newblock in {\em Proceedings of the 2012 IEEE International Conference on
  Acoustics, Speech and Signal Processing (ICASSP)}, 2012, pp. 4341--4344.

\bibitem{doi:10.1109/ICASSP.2013.6639103}
Po-Sen Huang, Kshitiz Kumar, Chaojun Liu, Yifan Gong, and Li~Deng,
\newblock ``{Predicting speech recognition confidence using deep learning with
  word identity and score features},''
\newblock in {\em Proceedings of the 2013 IEEE International Conference on
  Acoustics, Speech and Signal Processing (ICASSP)}, 2013, pp. 7413--7417.

\bibitem{doi:10.1109/ICASSP.2014.6854012}
Yik-Cheung Tam, Yun Lei, Jing Zheng, and Wen Wang,
\newblock ``{ASR error detection using recurrent neural network language model
  and complementary ASR},''
\newblock in {\em Proceedings of the 2014 IEEE International Conference on
  Acoustics, Speech and Signal Processing (ICASSP)}, 2014, pp. 2312--2316.

\bibitem{Korenevsky2015}
Maxim~L. Korenevsky, Andrey~B. Smirnov, and Valentin~S. Mendelev,
\newblock ``{Prediction of Speech Recognition Accuracy for Utterance
  Classification},''
\newblock in {\em Proceedings of Interspeech 2015}, 2015, pp. 1275--1279.

\bibitem{doi:10.1109/ICASSP.2015.7178796}
Atsunori Ogawa and Takaaki Hori,
\newblock ``{ASR error detection and recognition rate estimation using deep
  bidirectional recurrent neural networks},''
\newblock in {\em Proceedings of the 2015 IEEE International Conference on
  Acoustics, Speech and Signal Processing (ICASSP)}, 2015, pp. 4370--4374.

\bibitem{doi:10.1109/AICCSA.2016.7945669}
Rahhal Errattahi, Asmaa~El Hannani, Hassan Ouahmane, and Thomas Hain,
\newblock ``{Automatic speech recognition errors detection using supervised
  learning techniques},''
\newblock in {\em Proceedings of the 2016 IEEE/ACS 13th International
  Conference of Computer Systems and Applications (AICCSA)}, 2016, pp. 1--6.

\bibitem{doi:10.1016/J.SPECOM.2017.02.009}
Atsunori Ogawa and Takaaki Hori,
\newblock ``{Error detection and accuracy estimation in automatic speech
  recognition using deep bidirectional recurrent neural networks},''
\newblock {\em Speech Communication}, vol. 89, pp. 70--83, 2017.

\bibitem{doi:10.1109/SLT.2018.8639602}
Rahhal Errattahi, Salil Deena, Asmaa~El Hannani, Hassan Ouahmane, and Thomas
  Hain,
\newblock ``{Improving ASR Error Detection with RNNLM Adaptation},''
\newblock in {\em Proceedings of the 2018 IEEE Spoken Language Technology
  Workshop (SLT)}, 2018, pp. 190--196.

\bibitem{doi:10.21437/Interspeech.2019-2125}
Qiang Huang and Thomas Hain,
\newblock ``{Detecting Mismatch Between Speech and Transcription Using
  Cross-Modal Attention},''
\newblock in {\em Proceedings of Interspeech 2019}, 2019, pp. 584--588.

\bibitem{Povey2011}
Daniel Povey, Arnab Ghoshal, Gilles Boulianne, Luk{\'{a}}{\v{s}} Burget,
  Ondřej Glembek, Nagendra Goel, Mirko Hannemann, Petr Motli{\v{c}}ek, Yanmin
  Qian, Petr Schwarz, Jan Silovsk{\'{y}}, Georg Stemmer, and Karel
  Vesel{\'{y}},
\newblock ``{The Kaldi Speech Recognition Toolkit},''
\newblock in {\em Proceedings of the IEEE 2011 Workshop on Automatic Speech
  Recognition and Understanding}, 2011.

\bibitem{doi:10.1109/ICASSP.2015.7178964}
Vassil Panayotov, Guoguo Chen, Daniel Povey, and Sanjeev Khudanpur,
\newblock ``{Librispeech: An ASR corpus based on public domain audio books},''
\newblock in {\em Proceedings of the 2015 IEEE International Conference on
  Acoustics, Speech and Signal Processing (ICASSP)}, 2015, pp. 5206--5210.

\bibitem{Hain2013}
Thomas Hain and Oscar Saz,
\newblock ``{Factored WSJ-CAM0 Speech Corpus},'' [Online]. Available:
  https://mini.dcs.shef.ac.uk/resources/wsjcam0/. [Accessed: 2019-10-16], 2013.

\bibitem{SiTEC}
SiTEC,
\newblock ``{Speech Corpora},'' [Online]. Available:
  http://sitec.or.kr/{\#}speechcopora. [Accessed: 2019-10-16].

\bibitem{NCSOFT}
NCSOFT,
\newblock ``{Blade {\&} Soul},'' [Online]. Available: http://bns.plaync.com/.
  [Accessed: 2019-10-16].

\end{thebibliography}

\end{document}